\documentclass[10pt]{article}
\usepackage[cp866]{inputenc}
\usepackage[russian]{babel}

\begin{document}


\newcommand{\N}{N\raise.7ex\hbox{\underline{$\circ $}}$\;$}

\begin{center}
{\bf

\thispagestyle{empty}

BELARUS NATIONAL ACADEMY OF SCIENCES

B.I. STEPANOV's INSTITUTE OF PHYSICS

}
\end{center}

\vspace{50mm}
\begin{center}

{\bf

A.A. Bogush, V.V. Kisel, N.G. Tokarevskaya, V.M. Red'kov\footnote{E-mail: redkov@dragon.bas-net.by }\\
\vspace{5mm}

ON EQUATIONS FOR A SPIN-2 PARTICLE \\ IN EXTERNAL GRAVITATIONAL FIELD}

\vspace{5mm}

\end{center}

\begin{quotation}

30-component, of the first order, equation for a spin 2 particle, equivalent to
the second order Pauli-Fierz
one, is generalized to presence of an external electromagnetic field as well as
a curved background space-time geometry.
The essential  property of the generally covariant wave equation
obtained  is that here from the very beginning, in accordance
with requirement of the Pauli-Fierz approach, a set of additional relations on 30-component
wave function for eliminating complementary spin 0 and spin 1 fields is present at the starting equation.

\end{quotation}

\newpage

\subsection*{Introduction}

The theory  of the massive spin-2 field has received much attention over the years
since  the initial construction of a a Lagrangian formulation by Fierz and Pauli
[1-2].
The original Fierz-Pauli theory  for spin  was second order in derivatives $\partial_{\alpha}$
(and involved scalar and tensor auxiliary fields).
It is  highly  satisfactory as long  as we restrict ourselves to  a free particle case.
However this approach turned out not to be so good at considering spin-2 theory in presence an external
electromagnetic field.  Federbush [3] showed that  to avoid  a loss of constrains problem . the minimal
coupling had to be  supplemented by a direct non-minimal to the electromagnetic field strength.
There followed a number of works on modification or  generalizations  of the Fierz-Pauli theory
(Rivers [4], Nath [5], Bhargava and Watanabe [6], Tait [7], Reilly [8]).
At the same time  interest in  general high-spin  fields  was generated by  the discovery of the now
well-known inconsistency problems of Johnson and Sudarshan [9] and  Velo and Zwanzinger [10].
In the course of  investigating  their acausality problems for other then 3/2, Velo-Zwanzinger
rediscovered  the spin-2 loss of constrains problem, but were not at first aware of  the non-minimal
 term solution of it.
 (Velo [11]) later made  a  thorough analysis of the external field  problem
for  the 'correct'  non-minimally coupled  spin-2 theory, showing  that it  too is acausal.

All the work mentioned  above dealt  with  a second-order formalism for the spin-2 theory. Much of the
confusion  which arose over this theory  could be traced to  the so-called "derivative ordering ambiguity
(Naglal [12]). This problem  can be avoided by working from the start with a first-order  formalism
(for example see Gel'fand et al [13]) and for which  the minimal coupling procedure is unambiguous..

The work by Fedorov [14] was likely to be the first one where consistent investigation of
the spin-2 theory in the framework of first-order theory was carried out in detail.
The 30-component wave equation  [14] referred to the so-called canonical basis,
transition from which to the more familiar tensor formulation is possible but laborious task and it was not done in
{14].
Subsequently the same 30-component theory  was rediscovered and  fundamentally elaborated
in tensor-based approach by a number of authors (Regge [15], Schwinger [16],  Chang [17], Hagen [18],
 Mathews et al [19], Cox [20]). Also a matrix  formalism for the spin-2 theory was developed
 (Fedorod, Bogush, Krylov, Kisel  [21-25]).

Concurrently else one theory for spin-2 particle was advanced that requires 50 field  components
(Adler [26],  Deser et al [27], Fedorov and Krylov  [28, 23], Cox [20.]). It appears to be more
complicated, however some evident  correlation between  the corresponding massless theory  and
the non-linear gravitational  equation is revealed (Fedorov [28]).

Possible connections between two variants of spin-2 theories  have
been investigated. Seemingly, the most  clarity was achieved by
Bogush and Kisel [25], who showed that  50-component equation in
presence of an external electromagnetic field can be reduced to 30-component  equation
with additional interaction  that must be interpreted as anomalous magnetic momentum term.
Though this work was done in the framework of matrix formulation of the spinor technique (which
make this paper difficult to follow)  the main result was emphasized  quite  distinctly.

In the present work we will return to a 30-component theory and  will investigate it in
presence of both electromagnetic and gravitational fields. Gravitational fields are assumed to be
described in terms of a  curved space-time background. We are  going to trace in detail how
a  generally covariant Fierz-Pauli theory
can be derived from the  generally covariant first-order wave equation
(in presence of $A_{\mu}(x), g_{\alpha \beta}(x)$ fields).
Our  chief aim is to elucidate  the mechanism of resolving  the problem of  degrees of  freedom
in the 30-component theory.  The  most significant  aspect of the theory under consideration
is that it starts from a certain Lagrangian.
This means  that a presupposed set of additional conditions, constrains on 30 field constituents,
 turns out to be incorporated into the model from the very beginning, in accordance with  the
 known Fierz-Pauli program [1,2].

\begin{center}
{\bf 1.  Description of the  massive spin-2 particle on the base of
first-order formalism }
\end{center}

Let us  start off   with  the  tensor equations [21]
 $$
 A  \partial^{a} \Phi _{a} = m  \Phi
, \eqno(1.1a)
$$
$$
C \partial_{a} \Phi  +   B
\partial^{b} \Phi_{ab} = m  \Phi_{a}  ,
\eqno(1.1b)
$$
$$
 E ( \partial^{k} \Phi_{kab}  +  \partial^{k} \Phi_{kba}  -
{1 \over 2} g_{ab}   \partial^{k} \Phi_{kn}^{\;\;\;\;n} )  +
 N ( \partial_{a} \Phi_{b}  +  \partial_{b} \Phi_{a}
-  {1 \over 2} g_{ab}   \partial^{k} \Phi_{k} ) = m \Phi_{ab}  ,
\eqno(1.1c)
$$
$$
F  [  \partial_{a} \Phi_{bc}
 - \partial_{b} \Phi_{ac} +  {1 \over 3}  (g_{bc}
\partial^{k} \Phi_{ak}  -   g_{ac}  \partial^{k} \Phi_{bk} )
] = m  \Phi _{abc} ,
\eqno(1.1d)
$$

\noindent where $A,B, ...$ are   some constants.
From eq. (1.1c) it  follows  restrictions on $\Phi_{ab}$:
$$
\Phi_{ab} = + \Phi_{ba}  ,  g^{ab} \Phi_{ab} = \Phi^{a}_{a} = 0  .
\eqno(1.2)
$$

\noindent
In turn,  eq. (1.1d) presupposes  antisymmetry of $\Phi _{abc}$ over  two first indices
$
\Phi _{abc}  = - \Phi _{bac} \; .
$
Also,  simplifying eq (1.1d) over indices $b$ and $c$ and remembering eq. (1.2), we get to
$$
\Phi_{ab}^{\;\;\;\;b} = 0  .
\eqno(1.3a)
$$

\noindent In addition,  eq.  (1.1d) leads to
$$
\Phi_{abc}  +  \Phi_{bca}  +  \Phi_{cab} = 0 , \;\;  \mbox{или} \;\;\;
\epsilon^{kabc}   \Phi_{abc}  = 0
\eqno(1.3b)
$$

\noindent that  means  impossibility  to reduce the 3-d rank tensor to a simpler form.

Thus,  the total number of independent  components in  the used tensor set
 $\Phi, \Phi_{a}, \Phi_{ab}, \Phi_{abc}$  equals  to 30:
$$
\Phi (x)  -   1  ,
\Phi_{a}    -   4  ,
\Phi_{ab}   -   (10 - 1) = 9  ,
\Phi_{abc}  -   (6 \times 4 - 4 -4 ) = 16  .
\eqno(1.4)
$$

Now  it should be noted that without loss in generality one may  set the constants
$A$ and  $F$ equal to 1; this involves elementary converting
$$
{\Phi_{abc} \over F } , EF , {\Phi \over A} , CA \\\qquad \mbox{into} \qquad
\Phi_{abc}  , E   , \Phi  , C
$$

\noindent respectively. Correspondingly, instead of eqs. (1.1) further we will consider these
$$
\partial^{a} \Phi _{a} = m  \Phi  ,
\eqno(1.5a)
$$
$$
C \partial_{a} \Phi  +   B  \partial^{b} \Phi_{ab} = m  \Phi_{a}  ,
\eqno(1.5b)
$$
$$
E  ( \partial^{k} \Phi_{kab}  +  \partial^{k} \Phi_{kba} )  +
 N ( \partial_{a} \Phi_{b}  +  \partial_{b} \Phi_{a} -
{1 \over 2} g_{ab}   \partial^{k} \Phi_{k} ) = m  \Phi_{ab}  ,
\eqno(1.5c)
$$
$$
\partial_{a} \Phi_{bc}  - \partial_{b} \Phi_{ac} +  {1 \over 3}
(g_{bc}  \partial^{k} \Phi_{ak}  -   g_{ac}  \partial^{k} \Phi_{bk} )  = m  \Phi _{abc}  .
\eqno(1.5d)
$$

\noindent  taking in   mind  the constrains
$$
\Phi_{ab} = + \Phi_{ba}  ,   \Phi^{a}_{a} = 0  ,
\Phi _{abc}  = -  \Phi _{bac}  ,
 \Phi_{ab}^{\;\;\;\;b} = 0   ,
\Phi_{abc}  +  \Phi_{bca}  +  \Phi_{cab} = 0  .
\eqno(1.6)
$$

The first step is to exclude the first and third rank tensors from eq. (1.5).
To this end, with  the use of (1.5b) one gets
$$
E  ( \partial^{k} \Phi_{kab}  +  \partial^{k} \Phi_{kba} )  =
$$
$$
= { - E \over m}  [  {4\over 3}  (  \partial_{a} \partial^{k} \Phi_{kb}  +
\partial_{b} \partial^{k} \Phi_{ka} )  -  2  \partial^{k} \partial_{k} \Phi_{ab}
-  {2 \over 3}  g_{ab}  \partial^{k} \partial^{n} \Phi_{kn} )  ] ,
$$
$$
N [ \partial_{a} \Phi_{b}  +  \partial_{b} \Phi_{a} -
{1 \over 2} g_{ab}   \partial^{k} \Phi_{k}  ] =
{N \over m}  [  2C  (  \partial_{a} \partial_{b}  -
 {1 \over 4} g_{ab}   \partial^{k}  \partial_{k}  )  \Phi  +
$$
$$
+  B  ( \partial_{a}\partial^{k}
\Phi_{kb}  +  \partial_{b} \partial^{k} \Phi_{ka} -
{1 \over 2} g_{ab}  \partial^{k} \partial^{n} \Phi_{kn}  )  ]  .
$$

\noindent
Taking these two relations into account, eq. (1.5c) reads  as
($\nabla^{2} = \partial^{k} \partial_{k}  $)
$$
2C N  (  \partial_{a} \partial_{b}  -     {1 \over 4}
 g_{ab}  \nabla^{2}   )  \Phi  +   2  E  \nabla^{2}
\Phi_{ab}  +
\eqno(1.7)
$$
$$
 +  ( N B  -  {4 \over 3}  E  )
 ( \partial_{a} \partial^{k}  \Phi_{kb}  +
\partial_{b} \partial^{k} \Phi_{ka} - {1 \over 2}   g_{ab}
\partial^{k} \partial^{n} \Phi_{kn}  )  =  m^{2}  \Phi_{ab}  .
$$

\noindent In turn, from eq. (1.5а), remembering  (1.5б),  one  gets to
$$
 C  \nabla^{2}   \Phi  + B  \partial^{k} \partial^{n} \Phi_{kn} = m^{2}  \Phi  .
\eqno(1.8)
$$

The two relationships (1.7) and (1.8) provide  us with wave equations describing the spin-2 particle in two-order
formalism.  Now one needs to make clear how they could refer to a corresponding  Fierz-Pauli equations.
With this aim in mind we will produce certain constrains on yet arbitrary parameters
  entering eqs.   (1.5).
  Acting  on  eq.  (1.7) by  $\partial^{b}$, one produces
$$
{3 \over 2 } CN    \partial_{a}  \nabla^{2}   \Phi    -
{1 \over 2}  ( N B  -  {4 \over 3}  E  )
\partial_{a} \partial^{k} \partial^{n} \Phi_{kn} +
  ( N B  +  {2 \over 3}  E  )   \nabla^{2}  \partial^{k} \Phi_{ka} =
m^{2} \partial^{k} \Phi_{ka}  .
\eqno(1.9)
$$

\noindent  Setting
$$
 N B  +  {2 \over 3}  E   = 0   ,
\eqno(1.10)
$$

\noindent eq.  (1.9) looks as
$$
\partial_{a}   ( {3 \over 2} C N   \nabla^{2}   \Phi
 -  E      \partial^{k} \partial^{n}  \Phi_{kn}   ) = m^{2} \partial^{k} \Phi_{ka}  ,
\eqno(1.11a)
$$

\noindent
In the same time, acting on eq.  (1.8) by  $\partial_{a}$, one produces
$$
\partial_{a}
 ( C  \nabla^{2}   \Phi +
   B    \partial^{k} \partial^{n}  \Phi_{kn} ) =   m^{2}   \partial_{a} \Phi  .
\eqno(1.11b)
$$

\noindent
Let us demand that the free parameters  obey
$$
C = \mu^{-1}   {3 \over 2} CN  ,  B = -  \mu^{-1}  E  .
\eqno(1.11c)
$$

\noindent Then eq.  (1.11b) gets to the form
$$
\partial_{a}   (  {3 \over 2} CN   \   \Phi
 -   E      \partial^{k} \partial^{n}  \Phi_{kn}   )
=  \mu  m^{2}  \partial_{a} \Phi    .
\eqno(1.11d)
$$

\noindent Because the left-hand sides of eqs.  (1.11a)  and  (1.11d) coincide, then
$$
\partial^{k} \Phi_{kl} =   \mu  \partial_{l} \Phi  .
\eqno(1.12a)
$$

\noindent Subjecting eq. (1.12a) to operation $\partial^{l}$, one produces
$$
\mu  \nabla^{2}  \Phi - \partial^{k} \partial^{l} \Phi_{kl}  = 0  .
\eqno(1.12b)
$$

\noindent Let us  write down here  eq. (1.8) in the  form
$$
-B  (  - {C \over B}  \nabla^{2}   \Phi  -
 \partial^{k} \partial^{n} \Phi_{kn} ) = m^{2}  \Phi  .
\eqno(1.12c)
$$

\noindent and  let us demand that the equality
$
\mu = - {C \over B}  .
$
holds. Then
from eqs. (1.12c) and (1.12a) it follows
$$
\Phi = 0  .
\eqno(1.13a)
$$

\noindent  Therefore  eq.(1.12a) reads as
$$
\partial^{k} \Phi_{ka} =  0  .
\eqno(1.13b)
$$

It remains to proof that all these  equations on  free parameters are consistent with each others.
They all  are
$$
N B  +  {2 \over 3}  E   = 0  , C = \mu^{-1}   {3 \over 2} C N  , B = - \mu^{-1}  E  ,
\mu = - {C \over B}  ,
\eqno(1.14a)
$$

\noindent and they give
$$
\mu = {3 \over 2} N  , \qquad C = E = -{3 \over 2} NB  .
\eqno(1.14b)
$$

\noindent
Taking solutions of (1.14b)  in the form
$$
C = E = {1/2}  ,  N = 1 ,   B = - {1 \over 3}   ,
\eqno(1.15)
$$

\noindent it is easy to make sure that the system (1.7)-(1.8)  coincides  with
the Fierz-Pauli equations [1,2]\footnote{
In the following we adopt just this choice (1.15),  however
there are possible others. In particular, our choice differs from
that accepted in  [20,21]}
$$
(  \partial_{a} \partial_{b}  -
{1 \over 4} g_{ab} \nabla^{2}   )  \Phi   +
  ( \nabla^{2}   + M^{2} ) \Phi_{ab}  -
\eqno(1.16a)
$$
$$
-  ( \partial_{a} \partial^{k}  \Phi_{kb}  +  \partial_{b} \partial^{k} \Phi_{ka} -
{1 \over 2}   g_{ab}    \partial^{k} \partial^{n} \Phi_{kn}  )  =  0    ,
$$
$$
({1\over 2}  \nabla^{2}  + M^{2} )  \Phi  - { 1 \over 3}  \partial^{k} \partial^{n}
\Phi_{kn} =  0  .
\eqno(1.16b)
$$

\noindent
Here,  $m = i M$ and  $M$  is a real-valued massive parameter.

On taking into account the relations  (1.13a) and (1.13b) eqs.  (1.16) give
$$
(\nabla^{2} - M ^{2}) \Phi_{ab} = 0 \; ,
\;\;\; \Phi_{ab} = \Phi_{ba} \, ,  \Phi^{a}_{\;\;a} = 0  ,  \partial^{k} \Phi_{ka} = 0  ,
\eqno(1.17)
$$

\noindent these are  equations describing a free massive spin-2 particle [1,2].

Finally, let us write down the  initial first-order  system (1.15),
now with the fixed numerical parameters
$$
\partial^{a} \Phi _{a} = m  \Phi  ,
\eqno(1.18a)
$$
$$
{1 \over 2}  \partial_{a} \Phi  -  {1 \over 3}  \partial^{b} \Phi_{ab} = m  \Phi_{a}  ,
\eqno(1.18b)
$$
$$
{1 \over 2}
( \partial^{k} \Phi_{kab}  +  \partial^{k} \Phi_{kba}
-  {1 \over 2} g_{ab} \partial^{k} \Phi_{kn}^{\;\;\;\;n}\; )  +
\partial_{a} \Phi_{b}  +  \partial_{b} \Phi_{a} -
{1 \over 2} g_{ab}   \partial^{k} \Phi_{k}  = m  \Phi_{ab}  ,
\eqno(1.18c)
$$
$$
\partial_{a} \Phi_{bc}  - \partial_{b} \Phi_{ac} +  {1 \over 3}
(g_{bc}  \partial^{k} \Phi_{ak}  -   g_{ac}  \partial^{k} \Phi_{bk} )  = m  \Phi _{abc}  .
\eqno(1.18d)
$$

\noindent Further we will consider these equations as a basic ones  to describe
the massive spin-2 particle in Minkowsky space-time.
In its context the   Fierz-Pauli theory's status  should be revised and  their equations
should be regarded as derivative. And just eqs. (1.18) are to be extended to general relativity
case and investigated in that background.

\begin{center}
{\bf 2. Spin-2 particle in a curved space-time}
\end{center}

To take into account the presence of an external electromagnetic field and
a curved space-time background, , eqs. (1.18) are to be replaced by
$$
D^{\alpha} \Phi _{\alpha} = m  \Phi  ,
\eqno(2.1a)
$$
$$
{1 \over 2}  D_{\alpha} \Phi  -  {1 \over 3}
D^{\beta} \Phi_{\alpha \beta  } = m  \Phi_{\alpha}  ,
\eqno(2.1b)
$$
$$
{1 \over 2}
( D^{\rho } \Phi_{\rho  \alpha \beta }  +
D^{\rho } \Phi_{\rho  \beta  \alpha }
-   {1 \over 2} g_{\alpha  \beta}(x)   D^{\rho} \Phi_{\rho  \sigma}^{\;\;\;\;\sigma }
)  +
$$
$$
+   (  D_{\alpha} \Phi_{\beta}  +   D_{\beta} \Phi_{\alpha} -
{1 \over 2} g_{\alpha \beta }(x)   D^{\rho } \Phi_{\rho} )  =
m  \Phi_{\alpha \beta }  ,
\eqno(2.1c)
$$
$$
D_{\alpha} \Phi_{\beta\sigma }  - D_{\beta} \Phi_{\alpha  \sigma}
  +  {1 \over 3}  (g_{\beta  \sigma  }(x)   D^{\rho} \Phi_{\alpha \rho  }  -   g_{\alpha \sigma}(x)
D^{\rho } \Phi_{\beta \rho} )  = m  \Phi _{\alpha \beta \sigma}
\eqno(2.1d)
$$

\noindent where
$D_{\alpha} = \nabla_{\alpha} \; + \;ie \; A_{\alpha} \; $; $A_{\alpha}$ is  an electromagnetic 4-vector;
$\nabla_{\alpha}$  stands for a  generally covariant derivative. Again, the system (2.1)
involves the following constrains:
$$
\Phi_{\alpha  \beta } = \Phi _{\beta \alpha }  ,
\Phi_{\alpha \beta \rho }  = - \Phi_{\beta \alpha \rho }  ,
\Phi^{\;\;\alpha } _{\alpha}  = 0  ,  \Phi^{\alpha }_{\;\;\beta  \alpha } = 0 ,
\Phi_{\alpha \beta  \rho} + \Phi_{\beta  \rho \alpha }   + \Phi _{\rho \alpha \beta} = 0 \; .
$$

It should be stressed that such a straightforward and formal generalization of the above equations does not mean
that  the  taking  of eq. (1.18) into  eq. (2.1)  is trivial step without any substantial
peculiarities.  The things are quite to the contrary.  In particular,  the important  question of degrees of freedom
in the model  appears to become much more intricate.

Now we are going  to consider this problem in some detail.
As a first step,  let us exclude the first and third rank tensors from eq.
(2.1c). To this end, manipulating with eqs. (2.1d) and (2.1b) one can produce
two relations:
$$
D^{\rho }  \Phi_{\rho \alpha \beta }  +   D^{\rho }  \Phi_{\rho  \beta \alpha }  =
 {1 \over m}   [- (   D^{\rho  }   D_{\alpha}  \Phi _{ \beta \rho  }  +
                     D^{\rho }    D_{\beta }  \Phi _{\alpha  \rho  }  -
2  D^{\rho  }  D_{\rho}   \Phi_{\alpha \beta} )  -
$$
$$
- {1 \over 3}  (     D_{\alpha }   D^{\sigma}   \Phi_{\beta  \sigma }
 +   D_{\beta }   D^{\sigma}  \Phi_{\alpha  \sigma }
-
2  g_{\alpha \beta }(x)  D^{\rho }   D^{\sigma}   \Phi_{\rho  \sigma} )  ]   ,
\eqno(2.2a)
$$
$$
D_{\alpha}  \Phi_{\beta}   +   D_{\beta}  \Phi_{\alpha} -  {1 \over 2}  g_{\alpha \beta}(x)
   D^{\rho }  \Phi_{\rho} =
 {1 \over m}  [  {1 \over 2}  (  D_{\alpha }   D_{\beta}   +  D_{\beta }  D_{\alpha}  -
{1 \over 2}  g_{\alpha \beta }  D^{\rho}  D_{\rho}  )   \Phi -
$$
$$
-   {1 \over 3}  (  D_{\alpha}   D^{\rho}  \Phi_{\beta \rho}  +   D_{\beta}  D^{\rho }
 \Phi_{\alpha \rho}  -
{1 \over 2}  g_{\alpha  \beta }(x)  D^{\rho }  D^{\sigma}  \Phi_{\rho \sigma}  ) ]  .
\eqno(2.2b)
$$

\noindent
With the use of  (2.2), eq.  (2.1c)  can be brought to the form
$$
- {1 \over 2} (   D^{\rho}   D_{\alpha}  \Phi _{\rho  \beta}  +
            D^{\rho}   D_{\beta } \Phi _{\rho \alpha } )  +
D^{\rho }  D_{\rho}  \Phi_{\alpha  \beta}   +
$$
$$
- {1 \over 6}
 (     D_{\alpha } D^{\sigma}  \Phi_{\beta \sigma}
  + D_{\beta }  D^{\sigma} \Phi_{\alpha  \sigma }  )
+  {1 \over 3}   g_{\alpha  \beta } D^{\rho}  D^{\sigma}
 \Phi_{\rho  \sigma }  +
 {1 \over 2}  ( D_{\alpha}
D_{\beta} +   D_{\beta} D_{\alpha}  ) \Phi  -  {1 \over 4} g_{\alpha
\beta}(x) D^{\rho}  D_{\rho}   \Phi -
 $$
 $$
  -   {1 \over 3}  (
D_{\alpha} D^{\rho}  \Phi_{\beta  \rho } +   D_{\beta} D^{\rho}  \Phi_{\alpha \rho}
)  +  {1 \over 6}  g_{\alpha \beta} (x) D^{\rho}  D^{\sigma} \Phi_{\rho \sigma }
= m^{2}  \Phi _{\alpha \beta}  ,
 $$

\noindent
from which,  on elementary calculating  one gets  to
$$
- {1 \over 2}
 (   D^{\rho}   D_{\alpha}  \Phi _{\rho \beta }
+  D^{\rho } D_{\beta }   \Phi _{\rho \alpha} )
-
 {1 \over 2}
 (     D_{\alpha}  D^{\rho}  \Phi_{\beta \rho}
 +   D_{\beta}  D^{\rho}  \Phi_{\alpha \rho}   )  +
$$
$$
 + {1 \over 2} g_{\alpha \beta }(x)  D^{\rho} D^{\sigma }  \Phi_{\rho \sigma}  +
  D^{\rho  }  D_{\rho}  \Phi_{\alpha  \beta}   +
  {1 \over 2}  (  D_{\alpha } D_{\beta }  +   D_{\beta }  D_{\alpha }  ) \Phi  -
{1 \over 4}  g_{\alpha \beta }(x)
D^{\rho } D_{\rho }   \Phi
 = m^{2}  \Phi _{\alpha  \beta }  .
\eqno(2.3)
$$

\noindent The first expression in (2.3) is readily taken to the form
$$
-{1 \over 2}  ( D^{\rho }    D_{\alpha } \Phi _{\rho \beta  }  +
D^{\rho } D_{\beta }  \Phi _{\rho \alpha } )  =
$$
$$
=  -{1 \over 2} ( D_{\alpha} D^{\rho}  \Phi _{\beta \rho } +  D_{\beta}   D^{\rho}
\Phi _{\alpha \rho })  -
{1 \over 2} ( [D^{\rho}, D_{\alpha } ]_{-}   \Phi_{\rho \beta }
 +
[ D^{\rho}, D_{\beta }]_{-}  \Phi _{\rho \alpha  } )  .
$$

\noindent Correspondingly, eq.  (2.3) will  look as
$$
 -  D_{\alpha}  D^{\rho}  \Phi_{\beta  \rho }
- D_{\beta }  D^{\rho } \Phi_{\alpha \rho }  -
 {1 \over 2} ( [D^{\rho }, D_{\alpha }]_{-} \Phi_{\rho \beta }   +
[ D^{\rho}, D_{\beta}]_{-}  \Phi _{\rho \alpha  } )  +
$$
$$
+ {1 \over 2} g_{\alpha \beta } D^{\rho}  D^{\sigma}  \Phi_{\rho \sigma }
+  D^{\rho  }  D_{\rho}  \Phi_{\alpha \beta }   +
 {1 \over 2}  (  D_{\alpha} D_{\beta}   +   D_{\beta}  D_{\alpha}  )  \Phi  -
{1 \over 4}  g_{\alpha \beta }(x) D^{\rho} D_{\rho} \Phi
 = m^{2} \Phi _{\alpha \beta }  .
\eqno(2.4a)
$$

In turn, excluding with the help of (2.1b)the vector field from  eq.  (5.1a), one can produce
$$
 {1 \over 2} D^{\rho}   D_{\rho}    \Phi  -
{1 \over 3}  D^{\rho} D^{\sigma} \Phi_{\rho \sigma }  = m^{2}  \Phi  \; .
\eqno(2.4b)
$$

These  (2.4a) and (2.4b) provides with generalizations of eqs. (1.16a,b).
Formally,  eqs. (2.4) differ from eqs. (1.16) in changing
$\partial_{b}
\Longrightarrow D_{\beta }$  and  in appearance additional terms owing to the
$[...,...]_{-}$-commutator construction.

Although eqs.  (2.4) are rather complicated, they exhibit some interesting properties.
To bring them out we need perform some special manipulation.
Let us act on eq.  (2.4a) by $D^{\beta}$-operator:
$$
-  D^{\beta} D_{\alpha} D^{\rho}  \Phi_{\beta \rho} -
D^{\beta } D_{\beta} D^{\rho}  \Phi_{\alpha \rho}  -
 {1 \over 2}  D^{\beta} (   [D^{\rho}, D_{\alpha}]_{-} \Phi_{rho \beta}
 +
[ D^{\rho}, D_{\beta}]_{-}  \Phi _{\rho \alpha } )   +
\eqno(2.5a)
$$
$$
+  {1 \over 2}  D_{\alpha }  D^{\rho}  D^{\sigma}  \Phi_{\rho  \sigma}
 +   D^{\beta } D^{\rho  }  D_{\rho }  \Phi_{\alpha  \beta }  +
  {1 \over 2}  D^{\beta } D_{\alpha } D_{\beta} \Phi   +
 {1 \over 2}   D^{\beta } D_{\beta}  D_{\alpha }   \Phi  -
 {1 \over 4}  D_{\alpha } D^{\rho } D_{\rho} \Phi
 = m^{2}  D^{\beta } \Phi _{\alpha \beta }  ,
$$

\noindent which can be led to the form
$$
-  [ D^{\beta }, D_{\alpha }]_{-}  D^{\rho}  \Phi_{\beta  \rho} - D_{\alpha }  D^{\beta} D^{\rho}
\Phi_{\beta \rho }
-
D^{\beta } D_{\beta }  D^{\rho } \Phi_{\alpha \rho } -
$$
$$
- {1 \over 2}  D^{\beta } ( [D^{\rho}, D_{\alpha}]_{-}  \Phi_{\rho \beta }
 +
[ D^{\rho }, D_{\beta }]_{-}  \Phi _{\rho \alpha  })
 +
 {1 \over 2}  D_{\alpha }  D^{\rho } D^{\sigma } \Phi_{\rho \sigma }
+  D^{\beta }   D^{\rho  } D_{\rho }  \Phi_{\alpha \beta }  +
$$
$$
+  {1 \over 2}   [D^{\beta },  D_{\alpha}]_{-}  D_{\beta } \Phi   +
{1 \over 2}   D_{\alpha }  D^{\beta } D_{\beta } \Phi   +
 {1 \over 2}   D_{\alpha}   D^{\beta} D_{\beta}  \Phi  +  {1 \over 2}
[D^{\beta} D_{\beta},  D_{\alpha}]_{-}   \Phi   -
$$
$$
- {1 \over 4}  D_{\alpha}
D^{\beta} D_{\beta}   \Phi
 = m^{2}  D^{\beta}  \Phi _{\alpha \beta}  .
\eqno(2.5b)
$$

\noindent
From (2.5a) after evident simplification it follows
$$
m^{2} D^{\beta}  \Phi _{\alpha \beta } = {1 \over 2}  D_{\alpha}
(  {3 \over 2 } D^{\rho } D_{\rho } \Phi  - D^{\beta} D^{\rho} \Phi_{\beta \rho } )  +
\Delta_{\alpha}  ,
\eqno(2.6)
$$

\noindent   with the notation
$$
\Delta_{\alpha} =
 -     [D^{\beta}, D_{\beta}]_{-}  D^{\rho}  \Phi_{\beta \rho}
+ [ D^{\beta} ,   D^{\rho } D_{\rho}]_{-} \Phi_{\alpha \beta } -
$$
$$
- {1 \over 2}  D^{\beta} ( [D^{\rho}, D_{\alpha}]_{-}  \Phi_{\rho \beta }
  +   [ D^{\rho}, D_{\beta}]_{-} \Phi _{\rho \alpha })
+
 {1 \over 2}     [D^{\beta},  D_{\alpha }]_{-}  D_{\beta} \Phi   +
 {1 \over 2}   [D^{\beta } D_{\beta },  D_{\alpha }]_{-}  \Phi   .
\eqno(2.7)
$$

\noindent Now let us act on eq.  (2.4b)by  $ 3\; D_{\alpha}$-operator:
$$
D_{\alpha} ( {3 \over 2}  D^{\rho}   D_{\rho}    \Phi
-  D^{\rho} D^{\sigma} \Phi_{\rho \sigma }  )    =
3  m^{2} D_{\alpha}  \Phi   .
\eqno(2.8)
$$

\noindent
Taking into account eq. (2.8) in eq. (2.6), the latter is brought to the form
$$
m^{2} D^{\beta}  \Phi _{\alpha \beta } = {3 \over 2}    m^{2}  D_{\alpha }  \Phi
 +\Delta_{\alpha} ,
\;\;
\mbox{or} \;\;
m^{2} (  {3 \over 2}  D_{\alpha } \Phi  -   D^{\beta }  \Phi _{\alpha \beta }) = - \Delta_{\alpha}  .
\eqno(2.9)
$$

\noindent
Now acting on eq. (2.0) by operator  ${1 \over 3} \; D^{\alpha}$, we will have
$$
m^{2} ( - {1 \over 3}  D^{\alpha } D^{\beta}  \Phi _{\alpha \beta } + {1 \over 2}
  D^{\alpha } D_{\alpha }  \Phi  ) = - {1 \over 3}  D^{\alpha }
  \Delta_{\alpha},
$$

\noindent from where having remembered  eq.  (2.4b) we  arrive at
$$
m^{4} \Phi  = -{1 \over 3}  \partial^{\alpha } \Delta_{\alpha} .
\eqno(2.10)
$$

In a free case, the vector  $\Delta_{\alpha}$ vanishes identically, therefore
from  (2.10), (2.9) it follows
$
\Delta_{a} = 0 ,  \Phi =  0  ,  \partial^{b}  \Phi _{a b} = 0  .
$
As emphasized above, these two  constrains enable us to simplify noticeably
the starting wave equations (1.18): in fact reducing it to the form (1.18).

The fact of simplicity of the final equation  (1.17) is significantly valued,
especially  at finding its  solutions,  but from the formal viewpoint much  more
important and substantial matter is that existing constrains
$$
\Phi =  0  , \Phi_{a}^{a}=0,
 \partial^{b}  \Phi _{a b} = 0
\eqno(2.11)
$$

\noindent can be regarded as just those conditions which will rule out all subsidiary
components in the set $\Phi_{ab}, \Phi$. Correspondingly, only five remaining constituents
should be regarded as responsible for  description of
five degrees of  freedom  $2s+1 = 5$  of the  $s=2$ particle.

In this line of arguments the   generalized constrains on tensors  $\Phi,\Phi_{\alpha \beta} $
$$
\Phi_{\alpha}^{\alpha} = 0 , \qquad
m^{4} \Phi  = -{1 \over 3}  D^{\alpha} \Delta_{\alpha} , \qquad
m^{2} D^{\beta }  \Phi _{\alpha \beta } = {3 \over 2}    m^{2}  D_{\alpha }  \Phi
 + \Delta_{\alpha }  ,
\eqno(2.12)
$$

\noindent make the same work as  relations (2.11): they exclude subsidiary  field constituents.

Thus,
the extended theory of spin-2 particle in presence of external curved space-time background
 looks satisfactory in the line of arguments of    degrees of freedom.


\begin{center}
{\bf References}
\end{center}

\noindent
1. Pauli W., Fierz M. Uber relativistische  Feldgleichungen von Teilchen mit beliebigem Spin im
electro\-magnetishen Feld.  Helv. Phys. Acta. 1939, Bd. 12, S. 297-300.

\noindent
2.  Fierz V., Pauli W. On relativistic wave equations for particles of arbitrary spin in an
electromagnetic  field.  Proc. Roy. Soc. London. 1939, Vol. A173, P. 211-232.

\noindent
3.  Federbush P.  Nuovo Cimento. 1961. Vol. 19. P.572.

\noindent
4. Rivers R.J.   Nuovo Cimento. 1964. Vol.  34. P. 386.

\noindent
5. Nath L.M.   Nucl. Phys. 1965. Vol.  68. P. 600.

\noindent
6.
Bhargava S.C., Watanabe H. Nucl. Phys.  1966. Vol. 87. P.  273.

\noindent
7. Tait W.  Phys. Rev. D. 1972. Vol. 12. P. 3272.

\noindent
8. Reilly J.F.  Nucl. Phys. B. 1974. Vol. 70. P. 356.

\noindent
9. Johnson K., Sudarshan E.C.G.  Ann. Phys. NY. 1961. Vol.  85. P.  126.

\noindent
10.
Velo G.,  Zwanzinger D. Phys. Rev. 1969a. Vol. 186. P. 1337.

\noindent
11. Velo G. Nucl. Phys. B. 1972. vol. 43. P. 389.

\noindent 12.
12. Nagpal A.K. Lett. Nuovo Cimento. 1973. Vol.  Vol. 8. P. 353.

\noindent
13. Gelfand I.M., Yaglom A.M. JETP. 1948. Vol. 1948. P. 703-733.

\noindent
14. Fedorov F.I. To the theory of a particle with spin 2 (in Russian).  Uchenye Zapiski BGU.
Ser. fiz.-mat. . 1951. No  12. P. 156-173.

\noindent
15.
Regge T.  Nuovo Cimento. 1957. Vol. 5. P. 325.

\noindent
16. Schwinger S. Phys. Rev. 1963. Vol. 130. P.  800.

\noindent
17. Chang S.J.  Phys. Rev.  1966.  Vol. 148. P.  1259.

\noindent
18.
Hagen C.R.  Phys. Rew. D.  1972. Vol. 6. P. 984.

\noindent
19.
Mattews P.M. et al. J. Math. Phys.  1980. Vol. 21. P. 1495.

\noindent
20.
Cox W. First-order  formulation of massive spin-2 field theories.
J. Phys. A.: Math. Gen. 1982. Vol 15. P. 253- 268.

\noindent
21.
Krylov B.V.,  Fedorov F.I. First order equation for a graviton (in Russian).
DAN SSSR. 1967. Vol. 11. No 8. P. 681-684.

\noindent
22. Bogus A.A., Krylov B.V., Fedorov F.I. On matices of equation for  a spin 2 particle
(in Russian).  Vesti AN BSSR. Ser. fiz.-mat.  1968. No 1. P. 74-81.

\noindent
23.
Krylov B.V. On equations of first order for a  graviton (in Russsian).
Vesti AN BSSR. Ser. fiz.-mat. 1972. No  6. P. 82-89.

\noindent
24.  Kisel V.V., On relativistic wave equations for a massive particle with spin 2 (in Russian).
Vesti AN BSSR. Ser. fiz.-mat. 1986. No 5. P. 94-99.

\noindent
25.  Bogush A.A., Kisel V.V.
On description of the anomalous   magnetic momentum of a massive particle with spin 2 in the theory of
relativistic wave equation (in Russian).
Izvestia Vuzov. Fizika. . 1988. No 3. P. 11-16.

\noindent
26.
Adler D.  Canad. J. Phys. 1966. 1966. Vol. 44. P. 289.

\noindent
27.
Deser S.,  Trubatch J., Trubatch S. Canad. J. Phys. 1966. Vol.  44. Vol.  1715.

\noindent
28. Fedorov F.I. First order equations for a  graviton (in Russian). DAN SSSR. Mat.-Fiz.  1968.
Vol.  179. No 4. P.  802-805.

\end{document}